\documentclass[twocolumn,prd,preprintnumbers,amsmath,amssymb,nofootinbib]{revtex4}

\usepackage{graphicx}
\graphicspath{{FIGA_VGComment/}}
\usepackage{dcolumn}
\usepackage{bm}
\usepackage{epsfig}
\usepackage{epstopdf}
\usepackage{amssymb}
\usepackage{amsmath}
\usepackage{subfigure}
\usepackage{IEEEtrantools}
\usepackage{adjustbox}
\usepackage{color}
\usepackage{commath}
                              \newlength{\strikewidth}
                              \newlength{\strikelength}
\begin{document}
\setlength{\parskip}{1pt}

\title{Comment on ``Kinetic decoupling of WIMPs: Analytic expressions\textquotedblright}
\author{Isaac Raj Waldstein}
\email{isaac14@live.unc.edu}
\author{Adrienne L. Erickcek}
\affiliation{Department of Physics and Astronomy, University of North Carolina at Chapel Hill, Phillips Hall CB 3255, Chapel Hill, North Carolina 27599 USA}

%------------------------------------------------------------------------------

\begin{abstract}
Visinelli and Gondolo (2015, hereafter VG15) derived analytic expressions for the evolution of the dark matter temperature in a generic cosmological model. They then calculated the dark matter kinetic decoupling temperature $T_{\mathrm{kd}}$ and compared their results to the Gelmini and Gondolo (2008, hereafter GG08) calculation of $T_{\mathrm{kd}}$ in an early matter-dominated era (EMDE), which occurs when the Universe is dominated by either a decaying oscillating scalar field or a semistable massive particle before Big Bang nucleosynthesis. VG15 found that dark matter decouples at a lower temperature in an EMDE than it would in a radiation-dominated era, while GG08 found that dark matter decouples at a higher temperature in an EMDE than it would in a radiation-dominated era. VG15 attributed this discrepancy to the presence of a matching constant that ensures that the dark matter temperature is continuous during the transition from the EMDE to the subsequent radiation-dominated era and concluded that the GG08 result is incorrect.  We show that the disparity is due to the fact that VG15 compared $T_\mathrm{kd}$ in an EMDE to the decoupling temperature in a radiation-dominated universe that would result in the same dark matter temperature at late times.  Since decoupling during an EMDE leaves the dark matter colder than it would be if it decoupled during radiation domination, this temperature is much higher than $T_\mathrm{kd}$ in a standard thermal history, which is indeed lower than $T_{\mathrm{kd}}$ in an EMDE, as stated by GG08.
\end{abstract}

%\pacs{}

%\keywords{cosmology}

%------------------------------------------------------------------------------
% User-supplied List of keywords.
 
\maketitle         
%%%%%%%%%%%%%%%
\section{Introduction}
\label{sec:Intro}
%%%%%%%%%%%%%%%%       
In Ref.~\cite{Visinelli2015}, Visinelli and Gondolo (hereafter VG15), studied the kinetic decoupling of \mbox{dark matter (DM)} in various thermal histories. In particular, they considered an early matter-dominated era (EMDE), which occurs when the energy content of the Universe is dominated by either a decaying oscillating scalar field or a semistable massive particle prior to Big Bang nucleosynthesis (BBN). They defined the DM kinetic decoupling temperature $T_\mathrm{kd}$ in a general cosmology by
\begin{equation}\label{ratedef}
\gamma(T_\mathrm{kd}) = H(T_\mathrm{kd}),
\end{equation}   
where $\gamma$ is the momentum-transfer rate between DM and relativistic particles \cite{Visinelli2015, Hofmann2001} and $H$ is the expansion rate. In Sec. V of their article, VG15 compared the value of $T_\mathrm{kd}$ in an EMDE to the temperature at which DM would kinetically decouple from the plasma in a radiation-dominated (RD) era, $T_\mathrm{kd,std}$. VG15 found that the value of $T_\mathrm{kd}$ is less than the value of $T_\mathrm{kd,std}$, which they interpreted to mean that DM decouples at a lower temperature in an EMDE than it would in a RD era. 

This result contradicts Ref.~\cite{Gelmini2008}, hereafter GG08, which found that DM decouples at a higher temperature in an EMDE than it would in a RD era. \mbox{Figure 2 of Ref.~\cite{Visinelli2015}} highlights this apparent discrepancy for the case of \mbox{$p$\,-wave} scattering; it shows that \mbox{$T_{\mathrm{kd}}/T_{\mathrm{kd,std}} < 1$} for VG15 and \mbox{$T_{\mathrm{kd}}/T_{\mathrm{kd,std}} > 1$} for GG08. VG15 state that the presence of a matching constant that appears in their analytic expression for $T_\mathrm{kd}$ is responsible for their disagreement with GG08.  In this comment we show that the disparity is due to the way VG15 employs $T_\mathrm{kd,std}$ in their calculations. VG15 thought they were comparing the value of $T_\mathrm{kd}$ to the value of $T_\mathrm{kd,std}$ as defined by GG08: the temperature at which DM would decouple from the plasma in a RD era. However, VG15 used an expression for $T_\mathrm{kd,std}$ that corresponds to the temperature at which DM would have had to decouple in a RD era in order to reach the same temperature as it has at the end of an EMDE. This definition of $T_\mathrm{kd,std}$ only matches the GG08 definition if DM decouples during radiation domination. Since EMDE scenarios leave DM colder than it would be if it decoupled during radiation domination \cite{WEI2016, Gelmini2008}, the expression VG15 use for $T_\mathrm{kd,std}$ yields a much larger value of $T_\mathrm{kd,std}$ than the GG08 definition. We emphasize that the mistake that we have identified here does not affect the general expressions presented in VG15; it is confined to the usage of $T_\mathrm{kd,std}$ in the EMDE cosmologies considered in Sec. V of VG15.  
       
%%%%%%%%%%%%%%%
\section{Resolving The Discrepancy}
\label{sec: Resolve}
%%%%%%%%%%%%%%%%
The GG08 result in \mbox{Fig. 2 of VG15} is based on VG15 Eq.~(77),
\begin{equation}\label{eq77}
T_\mathrm{kd}^{\mathrm{GG}} =
\begin{cases}
\frac{T_\mathrm{kd,std}^{2}}{T_\mathrm{RH}},     & \text{for $T_\mathrm{kd,std} > T_\mathrm{RH}$},\\
T_\mathrm{kd,std},                                      &  \text{for $T_\mathrm{kd,std}< T_\mathrm{RH}$},
\end{cases} 
\end{equation}
where $T_\mathrm{kd,std}$ is given by Eq.~\eqref{ratedef} in a RD universe,
\begin{equation}\label{eqstar}
\gamma(T_\mathrm{kd,std}) = H^{\mathrm{rad}}(T_\mathrm{kd,std}).
\end{equation}
Here, $H^{\mathrm{rad}}(T)$ is the expansion rate in a RD universe as a function of the plasma temperature $T$. We will refer to the $T_\mathrm{kd,std}$ in Eq.~\eqref{eqstar} as $T_\mathrm{kd,std}^{\mathrm{R}}$, where the superscript ``R\textquotedblright \,stands for ``rate,\textquotedblright \,because it denotes the temperature at which the momentum-transfer rate falls below the Hubble rate. Thus, $T_\mathrm{kd,std}^{\mathrm{R}}$ is the temperature at which DM would decouple in a RD universe. If we use Eqs.~\eqref{ratedef} and \eqref{eqstar} to define $T_\mathrm{kd}$ and $T_\mathrm{kd,std}^{\mathrm{R}}$, respectively, then the GG08 conclusion that $T_\mathrm{kd}$ is greater than $T_\mathrm{kd,std}^{\mathrm{R}}$ must be correct. To see this, consider the Friedmann equation in a RD era: \mbox{$[H^\mathrm{rad}(T)]^{2} = (8\pi G/3) \rho_\mathrm{rad}(T)$}, where $G$ is the gravitational coupling constant and $\rho_\mathrm{rad}(T)$ is the energy density of radiation. If we consider the Hubble rate $H(T)$ in an EMDE at the same temperature $T$, then the Friedmann equation implies
\begin{equation}
H^{2}(T) =  \frac{8\pi G}{3} \left[ \rho_\mathrm{rad}(T) + \rho_{\phi}(T) \right] \gg [H^\mathrm{rad}(T)]^{2},
\end{equation}
because the energy density of the scalar field, $\rho_{\phi}(T)$, is much greater than $\rho_\mathrm{rad}(T)$ during the EMDE. Therefore,
\begin{equation}
\gamma(T_\mathrm{kd,std}^{\mathrm{R}}) = H^\mathrm{rad}(T_\mathrm{kd,std}^{\mathrm{R}}) < H(T_\mathrm{kd,std}^{\mathrm{R}}),
\end{equation}
which implies that at \mbox{$T=T_\mathrm{kd,std}^{\mathrm{R}}$} in an EMDE, DM \textit{has already decoupled} because the expansion rate exceeds the momentum-transfer rate. Therefore, DM decouples at a higher temperature (i.e. earlier) in an EMDE than it would in a RD era, as implied by the GG08 result.

How did VG15 arrive at the opposite conclusion---that DM decouples at a lower temperature in an EMDE than it would in a RD era? The answer is that VG15 effectively uses a different definition of $T_\mathrm{kd,std}$ than GG08, one that does not match Eq.~\eqref{eqstar} for $T_\mathrm{kd,std}^{\mathrm{R}}$ outside of a RD era. The VG15 result shown in \mbox{Fig. 2 of VG15} is based on their Eq.~(78),
\begin{equation}
T_{\mathrm{kd}}^\mathrm{VG} =
\begin{cases}\label{eq78}
\frac{T_\mathrm{kd, std}^{2}}{T_\mathrm{RH}}\,\left[1 + \frac{1}{\Gamma(3/4)}\,C_{2}\right]^{2}, & \text{for}\,\, T_{\mathrm{kd}} > T_{\mathrm{RH}},\\
\,\,T_\mathrm{kd, std}\,\left[1 + \frac{1}{\Gamma(3/4)}\,C_{2}\right], & \text{for}\,\,T_{\mathrm{kd}} < T_{\mathrm{RH}},
\end{cases}
\end{equation}
where $C_2$ is the value of the matching constant that ensures that the DM temperature is continuous during the transition from the EMDE to the ensuing RD era in VG15's ``broken power-law\textquotedblright \,cosmological model (see Sec. V of VG15), and $\Gamma(3/4)$ refers to the gamma function \mbox{$\Gamma(u) = \int_{0}^{\infty}\,dt\,e^{-t}\,t^{u-1}$}.  In their derivation of Eq.~\eqref{eq78}, VG15 used their Eq.~(57) to obtain the value of $T_\mathrm{kd,std}$.  This equation effectively defines a new standard kinetic decoupling temperature, which we will refer to as $T_\mathrm{kd, std}^{\mathrm{VG}}$: VG15 Eq.~(57) implies that 
\begin{equation}\label{eq57}
T_\mathrm{kd,std}^{\mathrm{VG}} = \frac{T^2(a_\mathrm{LT})}{T_{\chi}(a_\mathrm{LT})}\, \left( \frac{2}{2+n}\right)^{\frac{1}{2+n}}\, \Gamma\left(\frac{1+n}{2+n}\right),
\end{equation}
where $n\ge0$ is an integer set by the interactions between DM and the plasma, such that \mbox{$\gamma(T) \propto T^{4+n}$}. Equation~\eqref{eq57} must be evaluated long after the Universe becomes radiation dominated, such that the quantity $(T^{2}/T_{\chi})$ is constant. We denote the value of the scale factor at this time by $a_\mathrm{LT}$.    

The numerical factor
\begin{equation}
K(n) \equiv \left( \frac{2}{2+n}\right)^{\frac{1}{2+n}}\, \Gamma\left(\frac{1+n}{2+n}\right),  
\end{equation}
in Eq.~\eqref{eq57} was chosen to force Eq.~\eqref{eq57} to give the same value for the kinetic decoupling temperature as Eq.~\eqref{eqstar} if DM decouples in a RD era. While the value of $T_\mathrm{kd,std}^{\mathrm{VG}}$ equals the value of $T_\mathrm{kd,std}^{\mathrm{R}}$ if DM decouples during radiation domination, VG15 did not realize that the value of $T_\mathrm{kd,std}^{\mathrm{VG}}$ will be much greater than the value of $T_\mathrm{kd,std}^{\mathrm{R}}$ if DM decouples in an EMDE, which explains the tension between the VG15 and GG08 results. 

The reason for this temperature hierarchy is as follows: $T_\mathrm{kd,std}^{\mathrm{VG}}$ is the temperature at which DM would have had to decouple in a RD era in order to reach the same temperature as it has at the end of an EMDE. DM is much colder at the end of an EMDE than it would be if it decoupled during radiation domination \cite{WEI2016, Gelmini2008}. Therefore, $T_\mathrm{kd,std}^{\mathrm{VG}}$ must be much larger than $T_\mathrm{kd,std}^{\mathrm{R}}$, because DM would have had to decouple in a RD era much earlier than it normally would in order to end up as cold as it is at the end of an EMDE. To understand this effective definition of $T_\mathrm{kd,std}^{\mathrm{VG}}$ explicitly, consider \mbox{Fig. 1}, which compares the evolution of the plasma and DM temperatures in an EMDE scenario and a RD-only cosmology for \mbox{$p$\,-wave} scattering $(n=2)$. In the EMDE scenario, the Universe is dominated by a decaying oscillating scalar field until reheating, which occurs when the EMDE ends and radiation domination begins at \mbox{$a_\mathrm{RH} = 10^{8}$} and \mbox{$T_\mathrm{RH} = 5$ GeV}. In the RD-only cosmology shown in \mbox{Fig. 1}, the Universe is radiation dominated up to arbitrarily high plasma temperatures. In both of these cosmologies, the plasma temperatures share the same late-time behavior.
\begin{figure}[h]
\centering
\includegraphics[width=80mm,scale=20.5]{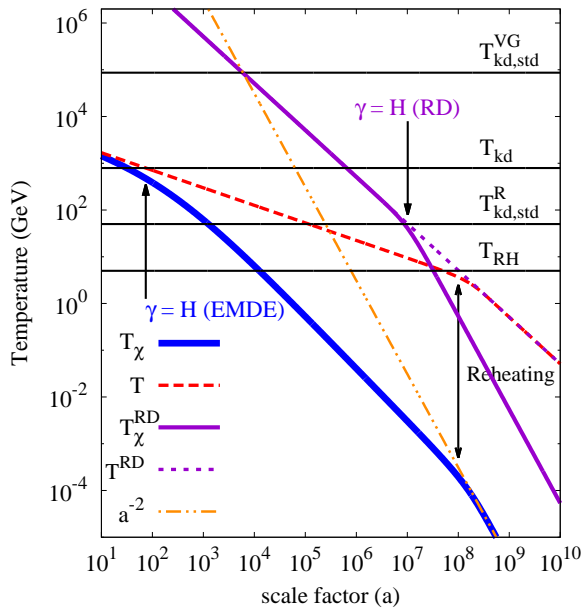}
\caption{The evolution of the plasma and DM temperatures in an EMDE scenario and a RD-only cosmology for \mbox{$p$\,-wave} scattering. The solid curves show the evolution of the DM temperature in the EMDE scenario ($T_{\chi}$) and in the RD-only cosmology ($T_{\chi}^{\mathrm{RD}}$). The dashed curves show the evolution of the plasma temperature in the EMDE ($T$) and in the RD-only cosmology ($T^{\mathrm{RD}}$). In the RD-only cosmology, the momentum-transfer rate $\gamma$ equals $H$ when \mbox{$T^{\mathrm{RD}} = T_\mathrm{kd,std}^{\mathrm{R}}= 50$\,GeV}. In the EMDE scenario, the Universe is dominated by a decaying oscillating scalar field until reheating, which occurs when the EMDE ends and radiation domination begins at \mbox{$a_\mathrm{RH} =10^{8}$} and \mbox{$T_\mathrm{RH}=5$\,GeV}. In the EMDE scenario, \mbox{$\gamma$ equals $H$} when \mbox{$T = T_\mathrm{kd} = 791$\,GeV}. The \mbox{$a^{-2}$ reference curve (dot-dot dashed)}, which is normalized to match $T_{\chi}$ at late times, and the \mbox{$T^{\mathrm{RD}}\propto a^{-1}$} curve, which is normalized to match $T$ at late times, intersect at \mbox{$a = 6077$}. Equation~\eqref{eq57} implies that this intersection point sets the value of $T_\mathrm{kd,std}^\mathrm{VG}/K(2)$, where \mbox{$K(2) \simeq 1.03$}. This figure shows that decoupling during the EMDE scenario requires the temperature hierarchy \mbox{$T_\mathrm{kd,std}^{\mathrm{VG}} > T_\mathrm{kd} > T_\mathrm{kd,std}^{\mathrm{R}} > T_\mathrm{RH}$.}}
\label{FIG: 1}
\end{figure} 

The dashed curves in \mbox{Fig. 1} depict the evolution of the plasma temperature in the EMDE scenario ($T$), and the evolution of the plasma temperature in the RD-only cosmology ($T^{\mathrm{RD}}$). Note that $T^{\mathrm{RD}}$ and $T$ are indistinguishable from each other after reheating, which certifies that \mbox{Fig. 1} is comparing two cosmologies for which the plasma temperatures share the same late-time behavior. Before reheating, \mbox{$T(a) \propto a^{-3/8}$} due to energy injection from the decaying scalar field \cite{Giudice2001}. 

The solid curves in \mbox{Fig. 1} show the evolution of the DM temperature in the EMDE cosmology ($T_{\chi}$), and the evolution of the DM temperature in the RD-only cosmology ($T_{\chi}^{\mathrm{RD}}$). In producing these curves we set \mbox{$T_{\mathrm{kd,std}}^{\mathrm{R}} = 50$ GeV}, which determines $\gamma(T)$ via Eq.~\eqref{eqstar} with $n=2$. Therefore, the EMDE and RD-only cosmologies in \mbox{Fig. 1} show DM with the same mass $m_{\chi}$ and velocity-averaged scattering cross section, \mbox{$\langle \sigma v \rangle\, \propto T^{2}$}. In the RD-only scenario \mbox{$T_{\chi}^{\mathrm{RD}}(a) \simeq T^{\mathrm{RD}}(a)$} while \mbox{$\gamma \gtrsim H$} and \mbox{$T_{\chi}^{\mathrm{RD}}(a) \propto a^{-2}$} while \mbox{$\gamma \lesssim H$}. In the EMDE scenario \mbox{$T_{\chi}(a)\simeq T(a)$} until \mbox{$\gamma \lesssim H$}, \mbox{$T_{\chi}(a) \propto a^{-9/8}$ (quasidecoupled \cite{WEI2016})} while \mbox{$\gamma \ll H$} during the EMDE, and \mbox{$T_{\chi}(a) \propto a^{-2}$} after reheating. 

\mbox{Figure 1} reveals why EMDE scenarios leave DM colder than it would be if it had decoupled during radiation domination \cite{WEI2016, Gelmini2008}. The presence of a dominant energy source other than radiation in EMDE scenarios requires $H(T)$ to be much greater than $H^{\mathrm{rad}}(T)$, which forces DM to decouple earlier than it would during radiation domination. As shown in the EMDE scenario of \mbox{Fig. 1}, earlier decoupling allows $T_{\chi}$ to start decaying faster than $a^{-1}$ much earlier than it does in the RD-only cosmology, which enables the DM to reach a much lower temperature at \mbox{$T = T_\mathrm{RH}$} than it reaches at the same plasma temperature in the RD-only scenario. 

To further understand why $T_{\chi}$ is much less than $T_{\chi}^{\mathrm{RD}}$ after the EMDE scenario in \mbox{Fig. 1}, consider the following: \mbox{Fig. 1} shows that $T_\mathrm{kd}$ is greater than $T_\mathrm{kd,std}^{\mathrm{R}}$ if DM decouples during the EMDE scenario, which is consistent with the GG08 relation in Eq.~\eqref{eq77}: \mbox{$T_\mathrm{kd} \simeq T_\mathrm{kd,std}^{2}/T_\mathrm{RH}$} for \mbox{$ T_\mathrm{kd,std} > T_\mathrm{RH}$} (see also Ref.~\cite{Erickcek2015long}). Since \mbox{$T \propto a^{-3/8}$} during the EMDE scenario and \mbox{$T_{\chi} \propto a^{-9/8}$} between decoupling and reheating \cite{WEI2016}, \mbox{$T_{\chi}(T_\mathrm{RH}) \simeq T_\mathrm{kd,std}\,(T_\mathrm{RH}/T_\mathrm{kd,std})^{5}$}. Therefore, the temperature of the DM particles is suppressed by a factor of $(T_\mathrm{RH}/T_\mathrm{kd,std})^{3}$ if they decouple during the EMDE scenario as opposed to if they decouple in the RD-only cosmology shown in \mbox{Fig. 1}. The two cosmologies in \mbox{Fig. 1} have identical plasma temperatures at late times, but they exhibit very different DM temperatures at late times. 

To apply the definition of $T_\mathrm{kd,std}^{\mathrm{VG}}$ in Eq.~\eqref{eq57} to the EMDE scenario shown in \mbox{Fig. 1}, we start at the solid \mbox{$T_{\chi}\propto a^{-2}$} segment and extrapolate back along the \mbox{dot-dot dashed $a^{-2}$} reference curve until we intersect the solid \mbox{$T^{\mathrm{RD}}\propto a^{-1}$} segment. The intersection fixes the value of $T_\mathrm{kd,std}^\mathrm{{VG}}/K(2)$, which implies that \mbox{$T_\mathrm{kd,std}^\mathrm{VG} = 86\,575$\, GeV}, as given by Eq.~\eqref{eq57}. Since $T_\mathrm{kd,std}^{\mathrm{VG}}$ far exceeds all of the other temperatures shown in \mbox{Fig. 1}, we have the temperature hierarchy \mbox{$T_\mathrm{kd,std}^\mathrm{{VG}} > T_\mathrm{kd} > T_\mathrm{kd,std}^\mathrm{{R}} > T_\mathrm{RH}$}, which confirms that the temperatures $T_\mathrm{kd,std}^{\mathrm{VG}}$ and $T_\mathrm{kd,std}^{\mathrm{R}}$ are very different quantities if DM decouples during the EMDE scenario. Whereas $T_\mathrm{kd,std}^{\mathrm{R}}$ is the temperature at which DM would decouple from the plasma in the RD-only cosmology, $T_\mathrm{kd,std}^{\mathrm{VG}}$ is the temperature at which DM would have had to decouple in the RD-only cosmology in order to reach the same temperature as it has after the EMDE. The only way for DM to end up as cold in the RD-only scenario as it is at the end of the EMDE is if it decouples in the RD-only scenario much earlier than it actually does. This is why the value of $T_\mathrm{kd,std}^{\mathrm{VG}}$ is so much larger than the value of $T_\mathrm{kd,std}^{\mathrm{R}}$ if DM decouples during the EMDE scenario.      

VG15 acknowledged the difference between the definitions of $T_\mathrm{kd,std}^\mathrm{VG}$ and $T_\mathrm{kd,std}^\mathrm{{R}}$, but they overlooked it when they specialized to $p$\,-wave scattering in an EMDE and compared Eq.~\eqref{eq78} to the GG08 relation in Eq.~\eqref{eq77}. As a result, VG15 concluded that \mbox{$T_\mathrm{kd} < T_\mathrm{kd,std}$} if DM decouples in an EMDE, in conflict with the GG08 result which states that \mbox{$T_\mathrm{kd} > T_\mathrm{kd,std}$} if DM decouples in an EMDE. VG15 incorrectly credited this tension to the matching constant $C_2$, but we now see that the tension is due to the difference between the effective definitions of $T_\mathrm{kd,std}$ in Eqs.~\eqref{eqstar} and \eqref{eq57}.

%%%%%%%%%%%%%%%
\section{Conclusion}
\label{sec: Conclusion}
%%%%%%%%%%%%%%%%
We have shown that the disagreement between the VG15 and GG08 expressions for $T_{\mathrm{kd}}/T_{\mathrm{kd,std}}$ is due to the fact that they effectively use different definitions of $T_\mathrm{kd,std}$. GG08 uses $T_\mathrm{kd,std}^{\mathrm{R}}$, the temperature at which DM would decouple in a RD era, while VG15 uses $T_\mathrm{kd,std}^{\mathrm{VG}}$, the temperature at which DM would have had to decouple in a RD era in order to attain the same temperature as it has at the end of an EMDE. As we illustrate in \mbox{Fig. 1}, these temperatures are vastly different if DM decouples in an EMDE scenario. The discrepancy between the VG15 and GG08 results is unrelated to the matching constant that links the EMDE to the subsequent RD era at reheating. In fact, there would have been no discrepancy if \mbox{Fig. 2 of VG15} had simply plotted \mbox{$T_\mathrm{kd}$ vs $T_\mathrm{RH}$}, instead of $T_{\mathrm{kd}}/T_{\mathrm{kd,std}}$  vs $T_{\mathrm{RH}}/T_{\mathrm{kd,std}}$, because both VG15 and GG08 define $T_\mathrm{kd}$ by Eq.~\eqref{ratedef}. 
 
%%%%%%%%%%%%%%%
\section*{ACKNOWLEDGMENTS}
\label{sec: Ack}
%%%%%%%%%%%%%%%%
The authors were partially supported by NSF Grant No. PHY-1417446. I.R.W. also acknowledges support from the Bahnson Fund at the University of North Carolina at Chapel Hill.\\

\end{document}